\documentclass{jpp}
\usepackage{graphicx}

\usepackage[utf8]{inputenc}
\usepackage[T1]{fontenc}
\usepackage{amsmath}
\usepackage{dcolumn}
\usepackage{bm}
\usepackage{placeins}
\usepackage{color}

\usepackage{subfigure}

\shorttitle{Intermediate VM equilibria}
\shortauthor{T. Neukirch, F. Wilson and O. Allanson}

\title{A family of Vlasov-Maxwell equilibrium distribution functions describing a transition from the Harris sheet to the force-free Harris sheet}

\author{T. Neukirch\aff{1}
\corresp{\email{tn3@st-andrews.ac.uk}},
F. Wilson\aff{1}
\and O. Allanson\aff{2}}

\affiliation{\aff{1}School of Mathematics and Statistics,
University of St Andrews, 
St Andrews, UK, KY16 9SS
\aff{2}Space and Atmospheric Electricity Group, 
Department of Meteorology, 
University of Reading, Reading, RG6 6BB, UK.}

\begin{document}

\maketitle

\begin{abstract}
We discuss a family of Vlasov-Maxwell equilibrium distribution functions for current sheet equilibria that are intermediate cases between the Harris sheet and the force-free (or modified) 
Harris sheet. These equilibrium distribution functions have potential applications to space and astrophysical plasmas.
The existence of these distribution function had been briefly discussed in by \citet{Harrison-2009b},
but here it is shown that their approach runs into problems in the limit where the guide field goes to zero. The nature of this problem will be discussed and 
an alternative approach will be suggested that avoids the problem. This is achieved by considering a
slight variation of the magnetic field profile, which allows a smooth transition between the Harris and force-free
Harris sheet cases.

\end{abstract}

\section{Introduction}

%
%
%
%

Current sheets are important for the structure and dynamical behaviour of many plasma systems. In space and astrophysical plasmas current sheets play a crucial role in magnetic activity processes by, for example, aiding 
the release of magnetic energy by magnetic reconnection. Current sheet equilibria are often used as starting points for
studying the dynamic behaviour of plasmas in, e.g., the solar atmosphere, the solar wind and planetary magnetospheres.

Many astrophysical plasmas can be described as collisionless and in this case the relevant equilibria are solutions of the steady-state Vlasov-Maxwell (VM) equations \citep[e.g.][]{Schindlerbook}. 
Since current sheets are strongly localised in space, they can often be well approximated by one-dimensional (1D) models \citep[see, e.g.][]{Roth-1996,Zelenyi-2011,Kocharovsky-2016,Neukirch-2017}.
An often used example of a 1D current sheet model is the Harris sheet \citep{Harris-1962}, which is a neutral sheet model that has been used extensively in studies of, e.g., magnetic reconnection \citep[e.g.][]{Kuznetsova-1998,Shay-1998,Hesse-1999,Kuznetsova-2000,Kuznetsova-2001,Hesse-2001b,Pritchett-2001,Rogers-2003,Hesse-2004,Ricci-2004,Pritchett-2004,Hesse-2005,Pritchett-2005,Hesse-2006,Daughton-2007,Wan-2008,Daughton-2011,Hesse-2011}. 

In some plasma systems, it can be more appropriate to use a current sheet model 
for which the pressure gradient is negligible.
Such models are termed \textit{force-free}, and satisfy the condition
$\textbf{j}\times\textbf{B}=0$, i.e. the current density and magnetic field are aligned with each other. 

The Harris sheet magnetic field is kept in force-balance by a pressure gradient, but one can also keep the system in a macroscopic force balance by adding a non-uniform guide field to the system
while the plasma pressure is constant.
The resulting configuration is often called the force-free Harris sheet. Equilibrium distribution functions for this configuration have been found, for example,
by \citet{Harrison-2009b, Neukirch-2009, Wilson-2011, Abraham-Shrauner-2013, Kolotkov-2015, Dorville-2015, Allanson-2015, Allanson-2016, Wilson-2017, Wilson-2018,Neukirch-2020} \citep[for further references on force-free 
Vlasov-Maxwell equilibria, see e.g.][]{Moratz-1966,Sestero-1967,Channell-1976,Correa-Restrepo-1993, Attico-1999,Bobrova-2001, Harrison-2009a, Vasko-2014a}. 

Similarly to the Harris sheet, collisionless force-free configurations have been used 
as initial conditions for particle-in-cell simulations of collisionless reconnection, using for example the exact equilibrium by \citet{Harrison-2009b} \citep[e.g.][]{Wilson-2016}, linear force-free equilibria \citep[e.g.][]{Bobrova-2001,Nishimura-2003,Bowers-2007} or
approximate force-free equilibria \citep[e.g.][]{Hesse-2005, Liu-2013, Guo-2014, Guo-2015, Zhou-2015, Guo-2016a, Guo-2016b, Fan-2016}.

In their paper, \citet{Harrison-2009b} also discussed the case of collisionless current sheets that are intermediate cases between the Harris sheet and the force-free Harris sheet, i.e. 
cases for which the macroscopic force-balance is provided by a combination of the plasma pressure gradient and the gradient of the magnetic pressure component provided by the non-uniform guide field. 
These equilibria and their DFs  self-consistently describe the transition from the Harris sheet to the force-free Harris sheet (or vice versa), but have so far not been studied in any detail.
Hence, in this paper we present an investigation of these collisionless current sheet equilibria. As this investigation will show, there are actually some problems with the DFs presented in
\citet{Harrison-2009b}, which limit their usefulness in practice. To circumvent these issues, we present a family of slightly modified magnetic field profiles and corresponding DFs which
avoid these problems, but still describe a transition between the Harris sheet and force-free Harris sheet as limiting cases.

We remark that in this paper we focus on a case in which the plasma temperature is uniform across the current sheet (i.e. an isothermal case). Distribution functions for non-isothermal force-free current sheets have been found by, e.g., \citet{Kolotkov-2015}, \citet{Wilson-2017}, and \citet{Neukirch-2020}. In principle the analysis carried out here could be generalised to 
these non-isothermal cases.
%
%
%
%

The paper is structured as follows; in Section \ref{sec:macro}, we briefly discuss the macroscopic equilibria of the Harris sheet, the force-free Harris sheet and
the intermediate cases. We then discuss the corresponding VM equilibrium distribution functions as given by \citet{Harrison-2009b} in Section \ref{sec:micro} and illustrate the problem associated with
the intermediate cases in the limit when the guide field amplitude tends to zero.
%
In Section \ref{sec: alternative}, we present a modified magnetic field model, 
which allows us to avoid these problems with the distribution function. We close with our summary and conclusions in Section \ref{sec: summary}.

\section{The macroscopic picture: Harris sheet, force-free Harris sheet and intermediate cases}
\label{sec:macro}




The macroscopic force balance for one-dimensional (1D) collisionless current sheet equilibria, with spatial variation only in the $z$-direction, is determined by \citep[e.g.][]{Mynick-1979a,Neukirch-2017}
\begin{equation}
\frac{d}{dz} \left[ \frac{B_x(z)^2+B_y(z)^2}{2 \mu_0} + P_{zz}(z) \right] = 0.
\label{eq:forcebalance}
\end{equation}
For collisionless equilibria, we are usually dealing with a pressure tensor and $P_{zz}$ is the only component of the pressure tensor which contributes to the force balance equation.

In this paper we focus on the family of equilibria defined by
\begin{eqnarray}
\mathbf{B}(z)&=&B_0\left(\tanh(z/L),  \frac{B_{y0}}{B_0 \cosh(z/L)}, 0\right),\label{eq:B_generalharris}\\
P_{zz}(z)&=& \frac{B_0^2 - B_{y0}^2}{2\mu_0 \cosh^2(z/L)}+P_{b}, \label{eq:Pzz_generalharris}
\end{eqnarray}
where $L$ represents the half-thickness of the current sheet, and $P_{b}\ge0$ is a constant background pressure. 
For completeness, we mention that the current density is given by
\begin{equation}
\mathbf{j}(z) = \frac{B_0}{\mu_0 L} \left(   \frac{B_{y0} \sinh(z/L)}{B_0 \cosh^2(z/L)},   \frac{1}{\cosh^2(z/L)}, 0  \right).
\label{eq:current_density}
\end{equation}

\begin{figure}
\centering\
\subfigure[]{\scalebox{0.4}{\includegraphics{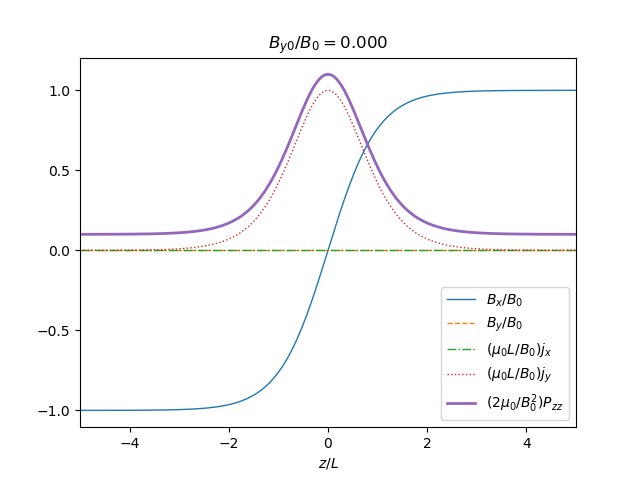}}}
\subfigure[]{\scalebox{0.4}{\includegraphics{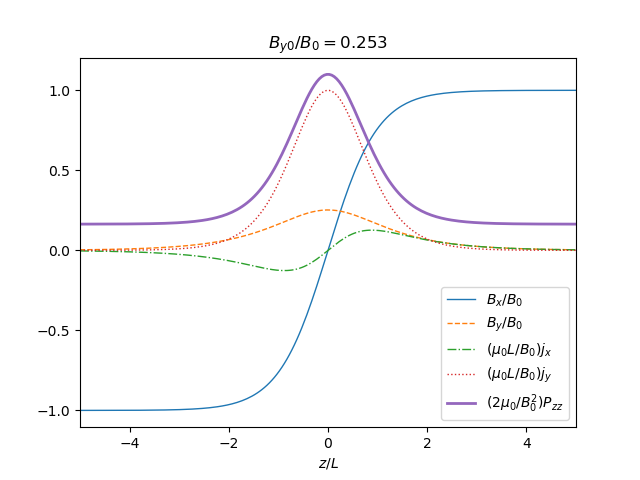}}}
\subfigure[]{\scalebox{0.4}{\includegraphics{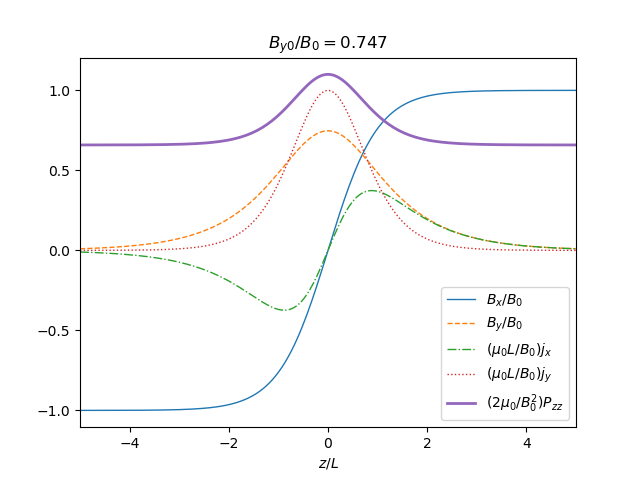}}}
\subfigure[]{\scalebox{0.4}{\includegraphics{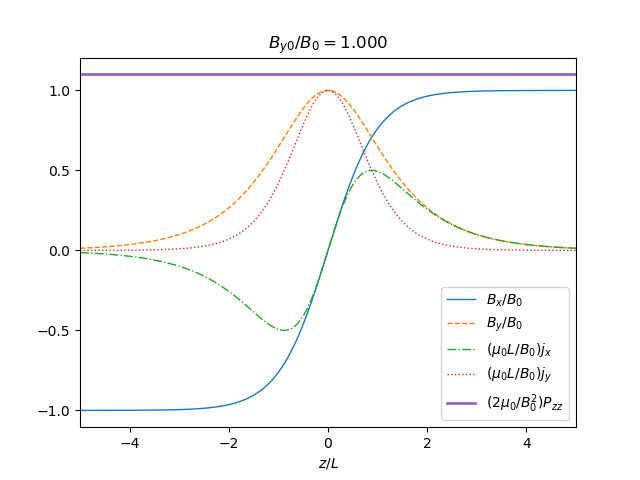}}}
\caption{Magnetic field, pressure, and current density profiles for (a) the Harris sheet, (b) and (c) intermediate cases with $B_{y0}/B_0\approx 0.25$ and $\approx 0.75$, respectively, and (d)  the force-free Harris sheet.
The background pressure $P_b$ (measured in units of $B_0^2/(2\mu_0)$) has been set to $B^2_{y0}/B_0^2 +0.1$ in each case.}
\label{fig:cs_profiles}
\end{figure}

The case $B_{y0} =0$ gives the Harris sheet \citep{Harris-1962}, which is a widely used 1D VM equilibrium in, e.g., reconnection studies, Often a constant guide field component is added to the Harris sheet field,
which is not included in our magnetic field model here.
%
%
When $B_{y0} = B_0$, we obtain the force-free Harris sheet, for which both the pressure $P_{zz} = P_b$ and the magnetic pressure $(B_x^2 + B_y^2)/2\mu_0 = B_0^2/2\mu_0$ are constant.
For $0< B_{y0} < B_0$ we get intermediate cases between the Harris sheet and the force-free Harris sheet.
Figure \ref{fig:cs_profiles} shows the magnetic field, pressure, and current density profiles for the Harris sheet, two intermediate cases, and force-free Harris sheet. 
Note that in the figure we have set the background pressure $P_b$ (measured in units of $B_0^2/(2\mu_0)$) to $B^2_{y0}/B_0^2 +0.1$.

At the macroscopic level discussed so far, there is no problem with varying $B_{y0}/B_0$ and in particular with letting this ratio go to $0$. This changes, however, when we consider the microscopic picture.

%

\section{The microscopic picture}
\label{sec:micro}

\subsection{1D Vlasov-Maxwell equilibria}

We assume a 1D Cartesian setup, in which all quantities depend only on the $z$-coordinate, and consider magnetic field profiles of the 
form $\textbf{B}=(B_x,B_y,0)$, for which $\textbf{B}=\nabla\times\textbf{A}$ (for vector potential $\textbf{A}=(A_x, A_y, 0)$). 
In this paper we will always impose conditions on the microscopic parameters of the DFs such that the electric potential $\phi$ (and hence the electric field) 
vanishes \citep[this can always be achieved for the cases we discuss here, see e.g.][]{Neukirch-2017}.
On the microscopic level, we assume that the distributions functions, $f_s$, are functions of the particle energy, 
$H_s=m_s(v_x^2+v_y^2+v_z^2)/2$, and the $x$- and $y$-components of the canonical momentum, $\textbf{p}_s=m_s\textbf{v}+q_s\textbf{A}$ (for $m_s$ the mass and $q_s$ 
the charge of species $s$, respectively), since these are known constants of motion for a time-independent system with spatial invariance in the $x$- and $y$-directions. 

Under the assumptions described above, the VM equations reduce to Amp\`{e}re's law in the form
\begin{eqnarray}
\frac{\mathrm{d}^2A_x}{\mathrm{d}z^2}&=&-\mu_0\frac{\partial P_{zz}}{\partial A_x}\label{ampx}\\
\frac{\mathrm{d}^2A_y}{\mathrm{d}z^2}&=&-\mu_0\frac{\partial P_{zz}}{\partial A_y}\label{ampy},
\end{eqnarray}
where $P_{zz}$ is  the only component of the pressure tensor that plays a role in the force-balance of the 1D equilibrium, defined by
\begin{equation}
P_{zz}(A_x, A_y)=\sum_s m_s\int v_z^2f_s(H_s, p_{xs}, p_{ys})\mathrm{d}^3v.\label{pzz def}
\end{equation} 
For a specified magnetic field profile, therefore, one needs to determine $P_{zz}(A_x, A_y)$ such that the vector potential associated with the given magnetic field is a solution of Amp\`{e}re's law.
Regarding Equation (\ref{pzz def}) as an integral equation for $f_s$ and solving it, will 
give DFs that self-consistently reproduce this macroscopic field profile\citep[e.g.][]{Channell-1976,Alpers-1969, Mottez-2003}. 
For some examples of the application of this approach, 
see \citet{Harrison-2009a, Harrison-2009b, Neukirch-2009, Wilson-2011, Abraham-Shrauner-2013, Kolotkov-2015, Allanson-2015, Allanson-2016, Wilson-2017, Wilson-2018}.

\subsection{The distribution functions}

\citet{Harrison-2009b} used Channell's method \citep{Channell-1976} to find the following DF for the force-free Harris sheet:
\begin{eqnarray}
f_s(H_s, p_{xs}, p_{ys})=\frac{n_{0s}}{\left(\sqrt{2\pi}v_{th,s}\right)^3}e^{-\beta_s H_s}\left[e^{\beta_s u_{ys}p_{ys}}+a_s\cos(\beta_su_{xs}p_{xs})+b_s\right].
\label{h&n df}
\end{eqnarray}
where $n_{0s}$ is a typical particle density for species $s$, $\beta_s = (k_\mathrm{b} T_s)^{-1}$ is the usual inverse temperature parameter and $v_{th,s}^2 = k_\mathrm{b} T_s/m_s = (m_s \beta_s)^{-1}$ is
the square of the thermal velocity of species $s$. As
discussed in detail in, for example, \citet{Neukirch-2009}, the additional parameters $a_s$, $b_s$, $u_{xs}$ and $u_{ys}$ have to satisfy further constraints to (i) have a positive DF ($b_s > a_s \ge 0$), (ii) 
guarantee that the electric potential vanishes, and (iii) ensure that the magnetic vector potential associated with the given macroscopic magnetic field is a solution of Amp\`{e}re's law.

The DF (\ref{h&n df}) is the sum of the Harris sheet DF \citep{Harris-1962} and an additional part, depending on $H_s$ and $p_{x,s}$
\begin{equation}
f_s(H_s, p_{xs}, p_{ys}) = f_{s, Harris}(H_s, p_{ys}) + \frac{n_{0s}}{\left(\sqrt{2\pi}v_{th,s}\right)^3} e^{-\beta_s H_s}\left(a_s\cos(\beta_su_{xs}p_{xs})+b_s\right),
\end{equation}
where the Harris sheet DF is given by
\begin{equation}
 f_{s, Harris}(H_s, p_{ys}) = \frac{n_{0s}}{\left(\sqrt{2\pi}v_{th,s}\right)^3}   e^{-\beta_s(H_s -u_{ys}p_{ys})}.
\end{equation}
\citet{Harrison-2009b} pointed out that by varying $a_s$ the DF (\ref{h&n df}) can in principle describe all the intermediate cases between the force-free and Harris cases and 
it looks as if in the limit $a_s\to0$ one should recover the Harris sheet DF. 

However, if one looks more carefully one finds that the parameter $a_s$ has to satisfy the relation \citep[similary to e.g.][]{Neukirch-2009}
\begin{equation}
a_{s}=\frac{B_{y0}^2}{2B_{0}^2}\exp\left(\frac{u_{xs}^2}{2v_{th,s}^2}\right) \exp\left(\frac{u_{ys}^2}{2v_{th,s}^2}\right),
\label{a_condition}
\end{equation}
where 
\begin{equation}
 u_{xs}^2  = \frac{4}{B_{y0}^2  \beta_s^2 q_s^2 L^2}.
\label{uxs_condition}
\end{equation}
This particular form for $a_s$ results from the consistency relations that the distribution functions have to satisfy for the electric potential calculated from the quasi-neutrality condition to vanish identically (this is a pre-requisite for being able to apply the method by \citet{Channell-1976}).
Hence, in the limit $B_{y0} \to 0$ $a_s$ does not go to zero, but to $\infty$, which is unacceptable for the distribution function. For finite $B_{y0}$, $a_s$ is finite but it increases rapidly 
as a function of $B_{y0}$. 

\begin{figure}
\centering\
\subfigure[]{\scalebox{0.4}{\includegraphics{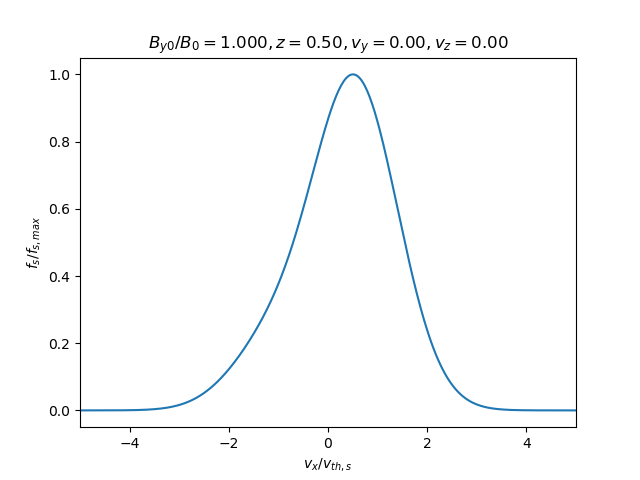}}}
\subfigure[]{\scalebox{0.4}{\includegraphics{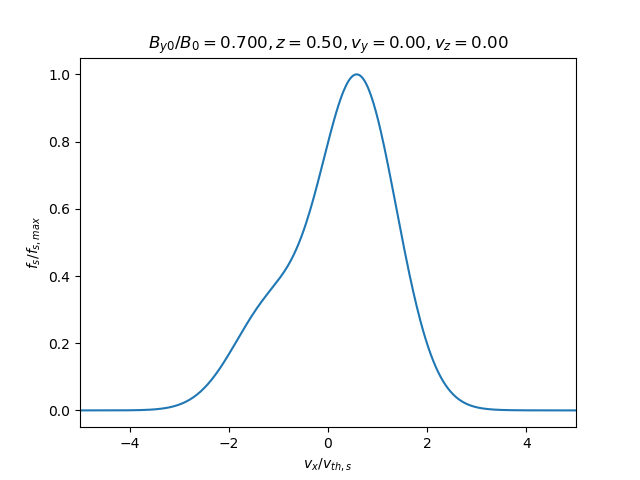}}}
\subfigure[]{\scalebox{0.4}{\includegraphics{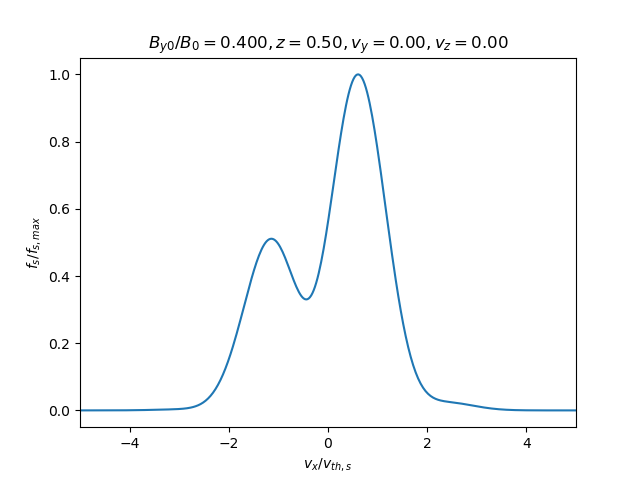}}}
\subfigure[]{\scalebox{0.4}{\includegraphics{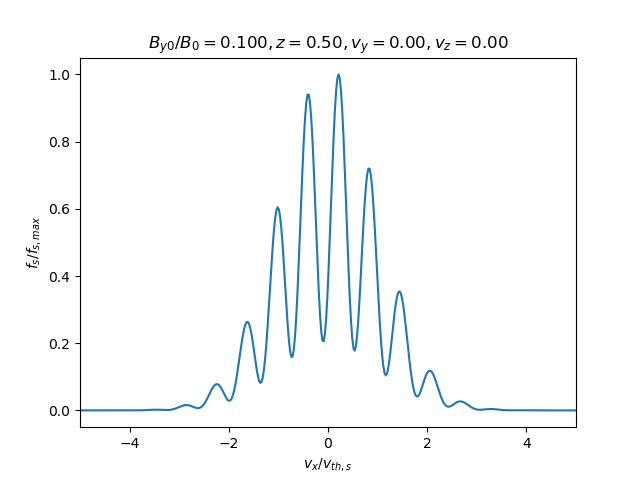}}}
\caption{Variation of the DF with $v_x/v_{th,s}$ for four different values of $B_{y0}/B_0$: panel (a) $B_{y0}/B_0 =1.0$, panel (b) $B_{y0}/B_0 =0.7$, panel (c) $B_{y0}/B_0 =0.4$, panel (d) $B_{y0}/B_0 =0.1$.
Here, each DF has been normalised by its maximum value and we have chosen $z/L=0.5$ and $v_y=v_z=0$.}
\label{fig:DF-intermediate}
\end{figure}

\begin{figure}
\centering\
\includegraphics[width=0.7\textwidth]{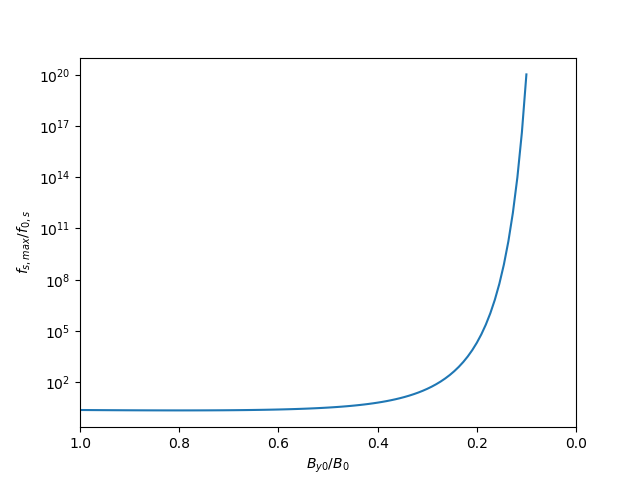}
\caption{Variation of the DF maximum with $B_{y0}/B_0$.}
\label{fig:amplitude-DF-intermediate}
\end{figure}

This leads to further unwanted properties of the DF, which we illustrate in Figures \ref{fig:DF-intermediate} and \ref{fig:amplitude-DF-intermediate}.
Figure \ref{fig:DF-intermediate} shows the DF as a function of $v_x/v_{th,s}$, for $z/L=0.5$ and
$v_y = v_z = 0$,
and how it changes as $B_{y0}/B_0$ decreases from $1.0$ to $0.1$. 
The maximum values of the DFs in Figure \ref{fig:DF-intermediate} are normalised to unity (i.e. we have divided the DFs by their maximum value for a given ratio $B_{y0}/B_0$). 
As one can clearly see in Figure \ref{fig:DF-intermediate} the DF develops more and more maxima and minima in the $v_x$-direction, due to the dominance of the cosine term in the distribution function caused by
the increase in $a_s$. It must be suspected that this filamentation in velocity space might lead to instabilities.
We also point out that the parameter $b_s$ has to increase as well so that $b_s > a_s$ to keep the DF positive (in the plots we have used $b_s = 1.5 \, a_s$). 
Figure \ref{fig:amplitude-DF-intermediate} shows on a logarithmic scale how the maximum value of the DF (for the fixed values of $z$, $v_y$ and $v_z$) increases dramatically as $B_{y0}/B_0$ decreases.

Taken together his clearly shows not only that the limit $B_{y0}/B_0 \to 0$ does not exist and that hence there is no smooth transition to the Harris sheet DF, but that the family of DFs will not be very useful even for finite,
but small values of the ratio $B_{y0}/B_0$. The question that arises is: can one find a family of VM equilibrium DFs which provides a smooth transition from the force-free Harris sheet to the Harris sheet?



\section{Alternative intermediate cases}
\label{sec: alternative}

In this section, we will consider an alternative magnetic field profile to that in Eq. (\ref{eq:B_generalharris}), of the form
\begin{equation}
\mathbf{B}(z)= B_0\left( \tanh(z/L),  \frac{\lambda}{ \cosh(\lambda z/L)}, 0\right),
\label{intermediate_field}
\end{equation}
where we have defined the abbreviation
\begin{equation}
\lambda = \frac{B_{y0}}{B_0}.
\end{equation}
A similar, albeit not totally identical magnetic field profile has previously been used by \citet{Huang-2017} to study instabilities using this type of collisionless current sheet. 
We will show that this magnetic field profile can be used to consistently describe a transition from the Harris sheet to the force-free Harris sheet. 

For completeness we here also state the current density and $zz$-component if the pressure tensor associated with this field, which are given by
\begin{eqnarray}
\mathbf{j}(z)&=&\frac{B_0}{\mu_0L}\left(\frac{\lambda^2 \sinh(\lambda z/L)}{\cosh^2(\lambda z/L)}, \frac{1}{\cosh^2(z/L)}, 0\right),\\
P_{zz}(z)&=&\frac{B_0^2}{2\mu_0}\left[  \frac{1}{\cosh^2(z/L)} + \lambda^2\left(1-\frac{1}{\cosh^2(\lambda z/L)}\right)\right]+P_{b2},
\end{eqnarray}
respectively, where $P_{b2}\ge0$ is a constant background pressure. We remark that we have written the non-background part of the pressure in such a way that it is always positive, 
regardless of the value of the positive constant $P_{b2}$. 

For $\lambda=0$, the magnetic field (\ref{intermediate_field}) becomes the Harris sheet field, and for $\lambda=1$ it becomes the 
force-free Harris sheet field. Setting $P_{b}= P_{b2} + B_0^2/2\mu_0$ in Eq. (\ref{eq:Pzz_generalharris}) for $\lambda=1$ ($B_{y0}=B_0$)
will make the two pressure functions equal in that case. 
The range $0<\lambda<1$ can be thought of as describing intermediate fields between the Harris and 
force-free Harris sheets, although the guide field and pressure profile deviates from the previous intermediate cases. 
\begin{figure}
\centering\
\subfigure[]{\scalebox{0.28}{\includegraphics{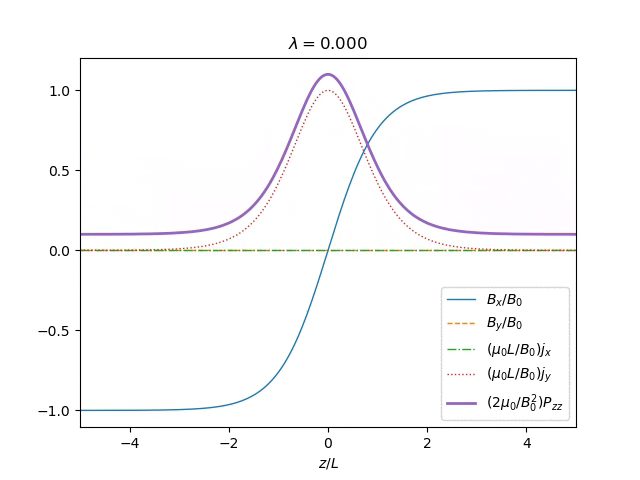}}}
\subfigure[]{\scalebox{0.28}{\includegraphics{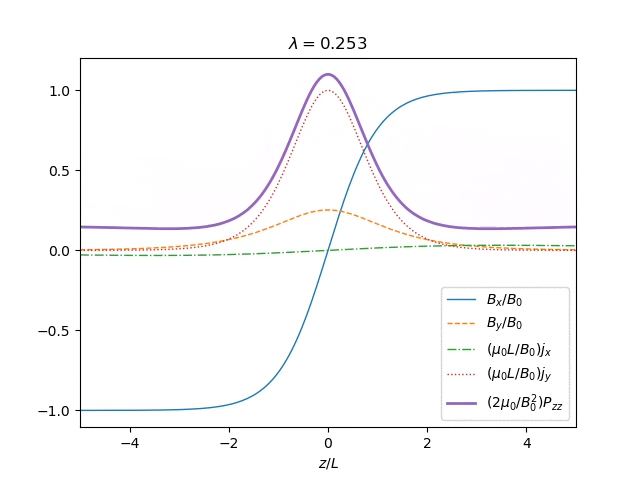}}}
\subfigure[]{\scalebox{0.28}{\includegraphics{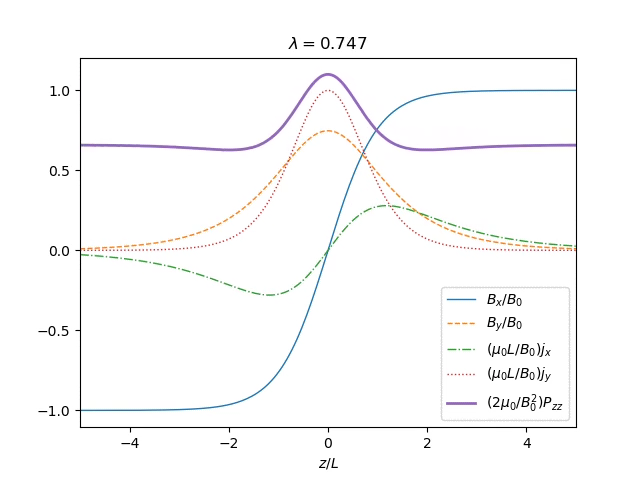}}}
\subfigure[]{\scalebox{0.28}{\includegraphics{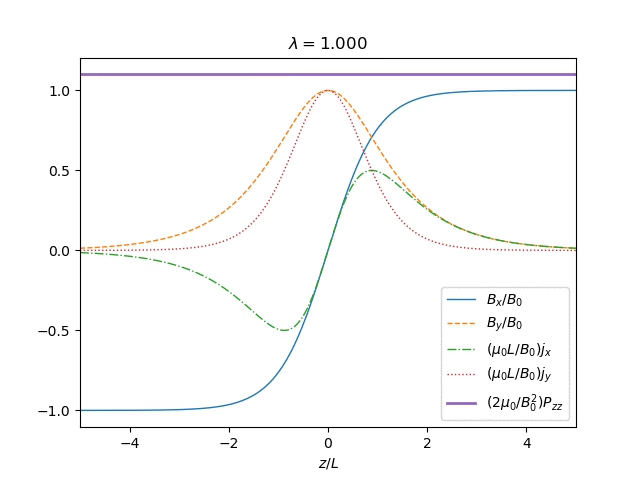}}}
\caption{Magnetic field, pressure, and current density for the alternative intermediate cases with (a)  $\lambda=0.0$, (b) $\lambda= 0.253$, (c) $\lambda = 0.747$ and (d) $\lambda = 1.0$.
The only noticeable change compared to the profiles shown in Figure \ref{fig:cs_profiles} are the slight dips (minima) in the pressure profile at the edge of the current sheet. The changes in the $B_y$ and
$j_x$ profiles are not immediately obvious without direct comparison.}
\label{fig:alt_int_profiles}
\end{figure}
Figure \ref{fig:alt_int_profiles} shows magnetic field, pressure and current density profiles for the Harris sheet ($\lambda=0.0$), two intermediate cases 
with $\lambda = 0.253$ and  $\lambda = 0.747$ and the force-free Harris sheet ($\lambda = 1.0$). The background pressure has been chosen
in the same way as for Figure \ref{fig:cs_profiles}. The only obvious difference to the plots shown in Figure \ref{fig:cs_profiles} are the slight dips 
in the pressure profile (local minima)
at the edges of the current sheet. On comparison with Figure \ref{fig:cs_profiles}, 
we also see that decreasing $\lambda$ results in a widening of the 
$B_y$ profile, due to the $\lambda$ factor inside the $\left[\cosh(\lambda z/L)\right]^{-1}$ in the $y$-component of Eq. (\ref{intermediate_field}). 
The amplitude of $B_y$ 
decreases in the same way as in the other case as $\lambda$ decreases, and eventually heads to zero as $\lambda\to0$.

It is straightforward to show that this macroscopic magnetic field profile is consistent with the DF in equation (\ref{h&n df}). One could suspect that this leads to the same problem with the limit $B_{y0} \to 0$ as before,
but when checking the constraints on the parameters of the DF one finds that while one still has
\begin{equation}
a_{s}= \lambda^2 \exp\left(\frac{u_{xs}^2}{2v_{th,s}^2}\right) \exp\left(\frac{u_{ys}^2}{2v_{th,s}^2}\right),
\label{a_condition_alternative}
\end{equation}
as before, the condition for $u_{xs}$ has changed to
\begin{equation}
 u_{xs}^2  = \frac{4 \lambda^2}{B_{y0}^2  \beta_s^2 q_s^2 L^2} = \frac{4 }{B_{0}^2  \beta_s^2 q_s^2 L^2},
\label{uxs_condition_alternative}
\end{equation}
which no longer varies with $B_{y0}$. Therefore, in this case we indeed find that $\lambda \to 0$ implies $a_s \to 0$, as desired.

\begin{figure}
\centering\
\subfigure[]{\scalebox{0.4}{\includegraphics{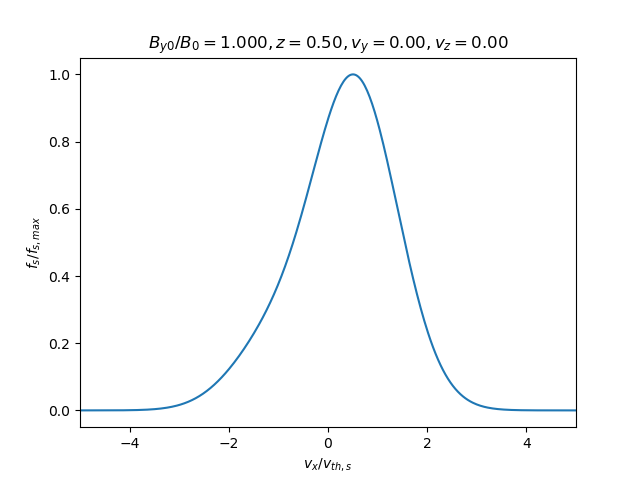}}}
\subfigure[]{\scalebox{0.4}{\includegraphics{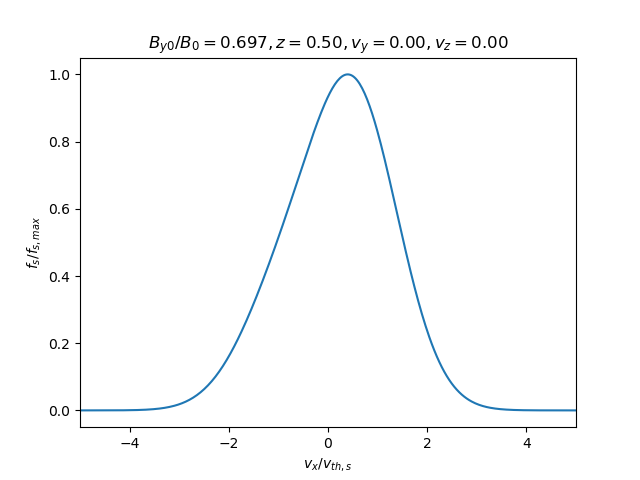}}}
\subfigure[]{\scalebox{0.4}{\includegraphics{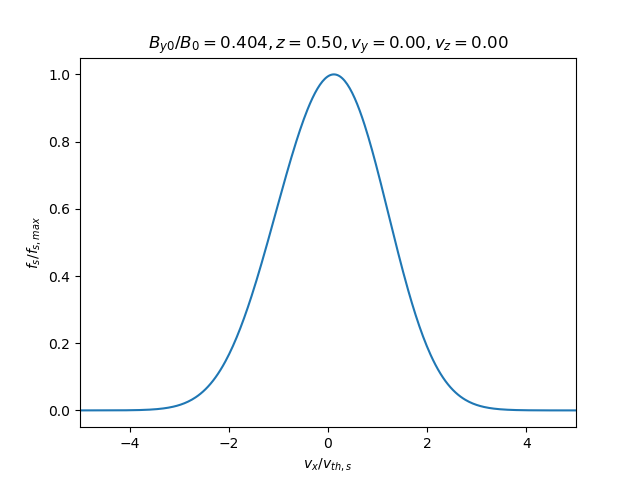}}}
\subfigure[]{\scalebox{0.4}{\includegraphics{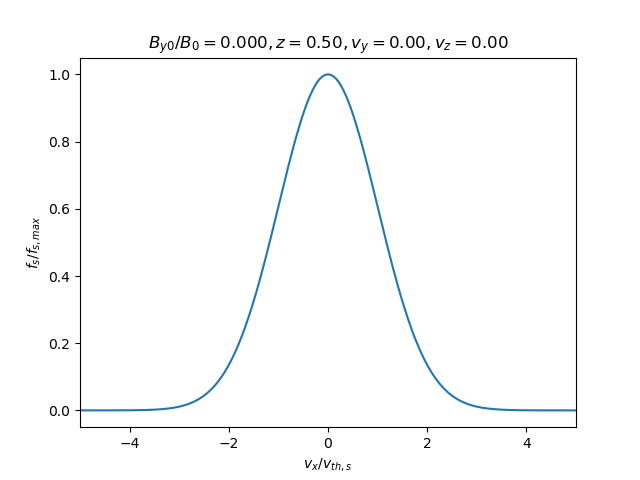}}}
\caption{Variation of the DF with $v_x/v_{th,s}$ for different values of $\lambda=B_{y0}/B_0$. Here, the DF has been normalised by its maximum value and we have chosen $z/L=0.5$ and $v_y=v_z=0$.
As one can see there is very little change as
$\lambda$ decreases.}
\label{fig:DF-intermediate_alt}
\end{figure}

\begin{figure}
\centering\
\includegraphics[width=0.7\textwidth]{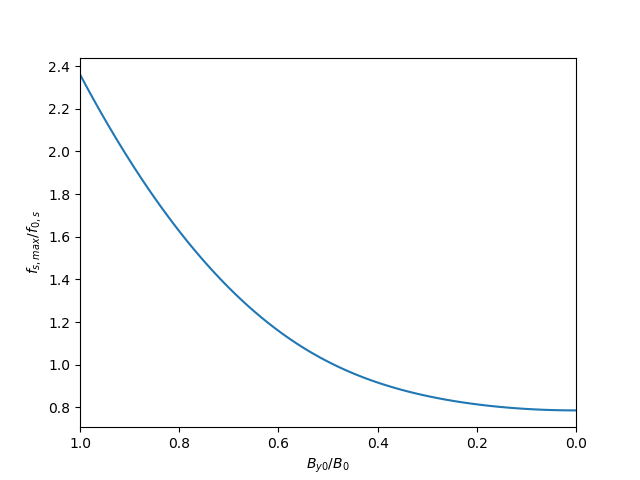}
\caption{Variation of the maximum of the DF with $B_{y0}/B_0$ for the alternative magnetic field profile. As one can see the maximum decreases with decreasing $\lambda$ and it does not diverge in the
limit $\lambda \to 0$. In contrast to Figure \ref{fig:amplitude-DF-intermediate} here a linear scale can be used for the plot.}
\label{fig:amplitude-DF-intermediate_alt}
\end{figure}

It is, however, prudent to also have a look at the DFs and their maximum value as $\lambda \to 0$. 
We show plots of the variation of the DF with $v_x$ (for $z/L=0.5$ and
$v_y = v_z = 0$) in Figure \ref{fig:DF-intermediate_alt}, for decreasing values of $\lambda$. For this case we can actually take the limit $\lambda \to 0$ without any problem (see panel (d)).
As in Figure \ref{fig:DF-intermediate}, we have normalised each DF to its maximum value. The variation of this maximum value with decreasing $\lambda$ is shown in 
Figure \ref{fig:amplitude-DF-intermediate_alt}. For this case the maximum of the DF actually decreases as $\lambda$ and hence $B_{y0}$ decreases.
With a 
relatively simple
modification of the magnetic field profile we have managed to eliminate the singular limit.

\section{Summary and Conclusions}
\label{sec: summary}

In this paper, we have discussed collisionless current sheet equilibria that are intermediate cases between the Harris sheet (current density perpendicular to magnetic field direction) and the force-free Harris sheet
(current density exactly parallel to the magnetic field direction).
Such a family of Vlasov-Maxwell equilibrium DF had already been briefly mentioned in \citet{Harrison-2009b}. However, as the more detailed investigation presented in this paper shows,
this family of DFs is of limited usefulness due to the fact that first of all the limit of the guide field amplitude $B_{y0} \to 0$ is singular in the sense that the maximum of the DF tends to $\infty$ and 
that with decreasing $B_{y0}$ the velocity space structure of the DF in the $v_x$ direction becomes more and more filamentary.
We proposed an alternative family of intermediate collisionless current sheet equilibria with a magnetic guide field that has a slightly modified spatial structure. Formally, the DFs associated with this magnetic field
remain the same, but the constraints imposed on the DF parameters by the self-consistency condition now allow the maximum value of the DFs to remain not only finite, but at a reasonable level as $B_{y0} \to 0$. 

We consider it important both from a theoretical and from a modelling/observational point-of-view that reasonable self-consistent equilibria of collisionless current sheets are available not only for the two limiting cases of force-free Harris sheet and normal Harris sheet. While some observations can be explained by, for example, force-free current sheet models \citep[e.g.][]{Panov-2011,Artemyev-2019a,Artemyev-2019b,Neukirch-2020}, 
it is to be expected that versions of the intermediate current sheet models are encountered with a greater likelihood than the limiting cases.\\

The authors acknowledge the support of the Science and Technology Facilities Council (STFC) via the consolidated grants ST/K000950/1, ST/N000609/1, and ST/S000402/1 (T.N. and F.W.) and the  Natural  Environment  Research  Council  (NERC) Highlight Topic Grant no.\ NE/P017274/1 (Rad-Sat) (O.A.). T.N. and  F.W. would also like to thank the University of St Andrews for general financial support.

\bibliographystyle{jpp}
\bibliography{final}

\end{document}